\documentstyle[amssymb,preprint,aps]{revtex}

\draft

\begin{document}
\title{Collisionless and hydrodynamic excitations of trapped boson-fermion mixtures}
\author{Xia-Ji Liu$^{1,2}$ and Hui Hu$^{3,4}$}
\address{$^1$European Laboratory for Nonlinear Spectroscopy and Dipartimento\\
diFisica, \\
Universit\`{a} di Firenze, Via Nello Carrara 1, 50019 Sesto Fiorentino, Italy%
\\
$^2$Institute of Theoretical Physics, Academia Sinica, Beijing 100080, China 
\\
$^3$Abdus Salam International Center for Theoretical Physics, P. O. Box 586, 
\\
Trieste 34100, Italy\\
$^4$Department of Physics, Tsinghua University, Beijing 100084, China}
\date{\today}
\maketitle

\begin{abstract}
Within a scaling ansatz formalism plus Thomas-Fermi approximation, we
investigate the collective excitations of a harmonically trapped
boson-fermion mixture in the collisionless and hydrodynamic limit at low
temperature. Both the monopole and quadrupole modes are considered in the
presence of spherical as well as cylindrically symmetric traps. In the
spherical traps, the frequency of monopole mode coincides in the
collisionless and hydrodynamic regime, suggesting that it might be undamped
in all collisional regimes. In contrast, for the quadrupole mode, the
frequency differs largely in these two limits. In particular, we find that
in the hydrodynamic regime the quadrupole oscillations with equal bosonic
and fermionic amplitudes generate an exact eigenstate of the system,
regardless of the boson-fermion interaction. This resembles the Kohn mode
for the dipole excitation. We discuss in some detail the behavior of
monopole and quadrupole modes as a function of boson-fermion coupling at
different boson-boson interaction strength. Analytic solutions valid at weak
and medium fermion-boson coupling are also derived and discussed.
\end{abstract}




\section{Introduction}

Shortly after the achievement of Bose-Einstein condensation of dilute,
magnetically trapped alkali atoms\cite{cornell,ketterle}, the investigation
of collective excitations in these systems has become a very active research
field (see \cite{review} for a recent theoretical review). The high accuracy
of frequency measurements and the sensitivity of collective phenomena to
interaction effects makes them good candidates to unravel the dynamical
correlation of the many-body system. So far, experimental results have been
obtained for low-lying collective modes of a trapped condensate in a wide
temperature regime, including breathing modes \cite{breathing}, surface
modes \cite{surface}, and the scissors mode \cite{scissors}. These
experiments have in turn stimulated a considerable amount of theoretical
work.

Recently the quantum degenerate regime has also been reached in a
magnetically trapped Fermi gas \cite{fermi}, and in a mixture of Bose and
Fermi particles \cite{mixture,roati}. The latter system is in particular
interesting since it serves as one typical example in which the intermingled
particles obey different statistics. Up to date the static property \cite
{molmer,amoruso,bijlsma,vichi}, the phase diagram and phase separation \cite
{nygaard,yi,viverit}, stability conditions \cite{roth,japan01} and
collective excitations \cite
{yip,valtan,pu,search,minguzzi,japan00,capuzzi,japan02} of trapped
boson-fermion mixtures have been theoretically investigated. In a recent
experiment, the collapse of a degenerated Fermi gas caused by the strong
attractive interaction with a Bose-Einstein condensate has been observed in
an atomic mixture of $^{40}${\rm K}$-^{87}${\rm Rb} \cite{modugno}, and
measurements of collective excitations might be available soon also in such
systems.

The purpose of the present paper is to study the collective excitations of
magnetically trapped boson-fermion binary mixtures in two different regimes:
a collisionless regime where the collision rate is small compared with the
frequencies of particle motion in the trap and a hydrodynamic (collisional)
regime in which collisions are sufficiently strong to ensure local
thermodynamic equilibrium. From the experimental point of view the
temperature $T$ of all the realized boson-fermion mixtures is around the
Fermi temperature $T_F$ (more precisely, $T\geqslant 0.2T_F$ \cite{mixture}%
), and the systems are possibly in or close to the hydrodynamic regime,
since the sympathetic cooling technique used in the experiments usually
requires a large boson-fermion interaction strength so that the frequent
collisions between fermions and bosons can foster the local thermal
equilibrium and ensure efficient thermalization of the fermionic component
to reach the quantum degeneracy. Of course, as far as the strongly
degenerate regime (where $T\ll T_F$) is concerned, the collisions are rare
because of Fermi statistic, and the systems will be finally in the
collisionless regime.

Several theoretical analysis have already been presented for the collective
excitations of a {\em spherically} trapped boson-fermion mixture. The
collisionless modes are considered by a sum-rule approach \cite{japan00} or
in the random-phase approximation \cite{capuzzi,japan02}. The collisional
collective oscillations are discussed by Minguzzi and Tosi \cite{minguzzi},
however, limited to the surface modes at weak fermion-boson coupling. On the
other hand, the {\em homogeneous} boson-fermion mixtures have also been
analytically studied \cite{yip,valtan,pu,search}. The repulsion between the
Bogoliubov phonon mode and zero-sound mode \cite{yip} (or Anderson mode \cite
{valtan}), is predicted when the degenerated Fermi gas is in the normal
collisionless limit (or in the superfluid phase).

In this paper, we shall analyze systematically the collisionless and
hydrodynamic monopole and quadrupole modes in the presence of spherical as
well as cylindrically symmetric traps \cite{note}. We discuss in some detail
the behavior of those modes against boson-fermion coupling at different
boson-boson interaction strength. In the spherical traps, we find that the
monopole frequency coincides in the collisionless regime and in the
hydrodynamic one, suggesting that it might be undamped in all collisional
regimes. In contrast, for the quadrupole mode the frequency differs
dramatically in these two limits. In particular, in the hydrodynamic regime
the quadrupole oscillations with equal bosonic and fermionic amplitudes are
found to generate an exact eigenstate of the system, resembling the Kohn
mode for the dipole excitation. Analytic solutions valid at weak and medium
fermion-boson coupling are also deduced and discussed.

The content of the paper is as follows. In the next section we derive
equations of the low-energy collective excitation of a boson-fermion mixture
in the Thomas-Fermi approximation by means of a scaling ansatz. In Sec. III
we first briefly describe the parameters and the numerical procedure
employed in the present calculation. We then turn to detailed discussion of
the results obtained by analyzing the dependence of the mode frequencies on
the boson-fermion coupling and boson-boson interaction strength, the
ground-state density distributions, and the mixing between bosonic and
fermionic collective oscillations. The last section is devoted to summary
and conclusions.

\section{Formulation}

We consider a dilute spin-polarized boson-fermion mixture trapped in a
cylindrically symmetric harmonic oscillator potential at low temperature. In
the semi-classical Thomas-Fermi approximation, the normal Fermi gas evolves
according to the Boltzmann-Vlasov kinetic equation \cite{odelin,menotti}
(see the Eq. (\ref{boltzmann}) below). In the collisionless regime, the
collisions are rare and we can safely neglect the collision term ($I_{coll}$%
) that accounts for the damping of collective modes. On the opposite of the
hydrodynamic regime, the collision term dominates and we resort to the Euler
equation of motion \cite{kagan}, which can be deduced from Boltzmann-Vlasov
kinetic equation under the assumption of local equilibrium for fermions \cite
{amoruso2}. For the bosonic part, we shall apply the same Stringari's
hydrodynamic formulation \cite{stringari} in both regimes, since the
dynamics of the condensate is less affected by the collisions \cite{remark}.

\subsection{collisionless regime}

According to the Stringari's hydrodynamic description \cite{stringari}, the
low-energy dynamics of the trapped bosonic atoms is determined by the
equations, 
\begin{eqnarray}
\frac{\partial n_b}{\partial t}+{\bf \nabla }\left( {\bf v}_bn_b\right) &=&0,
\nonumber \\
m_b\frac{\partial {\bf v}_b}{\partial t}+{\bf \nabla }\left(
V_{ho}^b+g_{bb}n_b+g_{bf}n_f+\frac 12m_b{\bf v}_b^2\right) &=&0,
\label{stringari}
\end{eqnarray}
where $V_{ho}^b\left( {\bf r}\right) =\frac 12m_b\left( \omega _{\bot
b}^2\rho ^2+\omega _{zb}^2z^2\right) $ is the cylindrical symmetric
confining potential, and $n_b\left( {\bf r},t\right) $, $n_f\left( {\bf r}%
,t\right) $ and ${\bf v}_b\left( {\bf r},t\right) $ are the boson, fermion
density and velocity field, respectively. The mean-field term $%
g_{bf}n_f\left( {\bf r},t\right) $ is included to take into account the
effect of boson-fermion interaction \cite{molmer}. The boson-boson and
boson-fermion interaction strength for the pseudopotentials, $g_{bb}$ and $%
g_{bf}$, are related to the $s$-wave scattering lengths $a_{bb}$ and $a_{bf}$
through $g_{bb}=4\pi \hbar ^2a_{bb}/m_b$, $g_{bf}=2\pi \hbar ^2a_{bf}/m_{bf}$
where $m_{bf}=m_bm_f/\left( m_b+m_f\right) $ is the reduced boson-fermion
mass. In Eq. (\ref{stringari}), we have already neglected the quantum
kinetic energy pressure term $\frac{\hbar ^2}{2m_b\sqrt{n_b}}{\bf \nabla }^2%
\sqrt{n_b}$ in the spirit of the Thomas-Fermi approximation \cite{review}.

For the fermionic part in the collisionless regime, in order to take into
account the effects of the boson-fermion interactions, we consider the
mean-field description based on the Boltzmann-Vlasov kinetic equation \cite
{odelin,menotti} without the collisional term ($I_{coll}$), 
\begin{equation}
\frac{\partial f}{\partial t}+{\bf v}_f\cdot \frac{\partial f}{\partial {\bf %
r}}-\frac 1{m_f}\frac{\partial V_{ho}^f}{\partial {\bf r}}\cdot \frac{%
\partial f}{\partial {\bf v}_f}-\frac{g_{bf}}{m_f}\frac{\partial n_b}{%
\partial {\bf r}}\cdot \frac{\partial f}{\partial {\bf v}_f}=0,
\label{boltzmann}
\end{equation}
where $f\left( {\bf r},{\bf v}_f,t\right) $ is the single particle phase
space distribution function for fermions, $n_f\left( {\bf r},t\right) =\int
d^3{\bf v}_ff\left( {\bf r},{\bf v}_f,t\right) $ and $V_{ho}^f\left( {\bf r}%
\right) =\frac 12m_f\left( \omega _{\bot f}^2\rho ^2+\omega
_{zf}^2z^2\right) $ is the confining potential. The last term in the left
hand side of Eq. (\ref{boltzmann}) is a Hartree-Fock mean-field term, also
known as the Vlasov contribution. The fermion-fermion interaction has been
neglected as the polarized system is considered \cite{butts}.

Without the boson-fermion interaction ($g_{bf}=0$), both the Eqs. (\ref
{stringari}) and (\ref{boltzmann}) admit the simple scaling solution, i.e., 
\begin{eqnarray}
n_b\left( {\bf r},t\right) &=&\frac 1{\prod_jb_j\left( t\right) }n_b^0\left( 
\frac{r_i}{b_i(t)}\right) ,  \nonumber \\
v_{bi}\left( {\bf r},t\right) &=&\frac 1{b_i(t)}\frac{db_i(t)}{dt}r_i,
\label{bansatz}
\end{eqnarray}
for bosons \cite{kagan,castin} and 
\begin{eqnarray}
f\left( {\bf r},{\bf v}_f,t\right) &=&f_0\left( \frac{r_i}{\gamma _i(t)},%
{\bf \tilde{v}}_f({\bf r},t)\right) ,  \nonumber \\
{\bf \tilde{v}}_{fi}({\bf r},t) &=&\gamma _i(t)v_{fi}-\frac{d\gamma _i(t)}{dt%
}r_i,  \label{fansatz}
\end{eqnarray}
for fermions \cite{odelin,menotti,bruun}. Here $n_b^0$ and $f_0$ are the
equilibrium distributions. The dependence on time $t$ is entirely contained
the six scaling parameters, $b_i\left( t\right) $ and $\gamma _i(t)$, where $%
i=x,$ $y,$and $z$. By substituting this solution into Eqs. (\ref{stringari})
and (\ref{boltzmann}), it is easily to show that, the scaling parameters
obey the coupled differential equations \cite{castin,bruun}, 
\begin{eqnarray}
\ddot{b}_i(t)+\omega _{ib}^2(t)b_i(t)-\frac{\omega _{ib}^2(0)}{%
b_i(t)\prod_jb_j\left( t\right) } &=&0,  \label{bfree} \\
\ddot{\gamma}_i(t)+\omega _{if}^2(t)\gamma _i(t)-\frac{\omega _{if}^2(0)}{%
\gamma _i^3(t)} &=&0.  \label{ffree}
\end{eqnarray}
Solutions of Eqs. (\ref{bfree}) and (\ref{ffree}) determine the evolution of
both the boson and fermion density. In particular, the eigenfrequencies of
small oscillations with $\omega _i(t)=\omega _i(0)$ are the resonance
frequencies of the collective density (shape) oscillations under the weak
perturbation of the external field. After linearizing around the equilibrium
values $b_i=1$ and $\gamma _i=1$, one finds, in the case of the spherical
harmonic traps, the result: $\omega _{Mb}=\sqrt{5}\omega _b$, $\omega _{Qb}=%
\sqrt{2}\omega _b$ and $\omega _{Mf}=2\omega _f$, $\omega _{Qf}=2\omega _f$
for the frequencies of the monopole and quadrupole oscillations,
respectively, which is already well known in the literatures.

In the presence of the boson-fermion interaction ($g_{bf}\neq 0$), however,
the simple scaling solution is no longer satisfied at every position ${\bf r}
$ after the substitution. A useful approximation, in the first order of $%
g_{bf}$, is to assume the scaling form of the solution as a priori, and
fulfill it on average by integrating over the spatial coordinates. The same
strategy has been recently used by Guery-Odelin \cite{odelin} to investigate
the effect of the interaction on the collective oscillation of a classical
gas in the collisionless regime and by Menotti {\it et al.} to study the
expansion of an interacting Fermi gas \cite{menotti}. In the latter, the
authors showed that the frequencies of the monopole and quadrupole modes for
isotropic traps deduced in this approximation coincide with the result
derived earlier by using the sum-rule approach. Of course, the assumption of
the scaling ansatz is only meaningful for the small value of $\left|
g_{bf}\right| $. It will apparently break down for a large and positive $%
g_{bf}$, at which the phase separation occurs.

In the approximation specified above, we substitute the scaling ansatz Eq. (%
\ref{bansatz}) into Stringari's hydrodynamic equations. By setting $%
R_i=r_i/b_i(t)$, one finds, 
\begin{eqnarray}
&&\ddot{b}_i(t)R_i+\omega _{ib}^2(t)b_i(t)R_i+\frac{g_{bb}}{m_b}\frac 1{%
b_i(t)\prod_jb_j\left( t\right) }\frac{\partial n_b^0\left( {\bf R}\right) }{%
\partial R_i}  \nonumber \\
&&\left. \hspace{0.8in}+\frac{g_{bf}}{m_b}\frac 1{b_i(t)\prod_j\gamma
_j\left( t\right) }\frac{\partial n_f^0(\frac{b_i\left( t\right) }{\gamma
_i\left( t\right) }R_i)}{\partial R_i}=0\right. ,  \label{btmp}
\end{eqnarray}
which reduces to the following form in equilibrium state, 
\begin{equation}
\omega _{ib}^2(t)R_i+\frac{g_{bb}}{m_b}\frac{\partial n_b^0\left( {\bf R}%
\right) }{\partial R_i}+\frac{g_{bf}}{m_b}\frac{\partial n_f^0({\bf R})}{%
\partial R_i}=0.  \label{btmp2}
\end{equation}
The equations for the scaling parameters $b_i(t)$ can be obtained by
multiplying Eq. (\ref{btmp}) by $R_in_b^0\left( R_i\right) $ on both sides
and integrating over the spatial coordinates. Making use of the equilibrium
properties of the density distribution (\ref{btmp2}), after some
straightforward algebra one finds, 
\begin{eqnarray}
&&\ddot{b}_i(t)+\omega _{ib}^2(t)b_i(t)-\frac{\omega _{ib}^2(0)}{%
b_i(t)\prod_jb_j\left( t\right) }  \nonumber \\
&&+\frac{g_{bf}}{m_bN_b\left\langle R_i^2\right\rangle _b}\frac 1{%
b_i\prod_jb_j}\int d^3{\bf R}\frac{\partial n_f^0({\bf R})}{\partial R_i}%
R_in_b^0(\frac{\gamma _i}{b_i}R_i)  \nonumber \\
&&\left. -\frac{g_{bf}}{m_bN_b\left\langle R_i^2\right\rangle _b}\frac 1{%
b_i\prod_jb_j}\int d^3{\bf R}\frac{\partial n_f^0({\bf R})}{\partial R_i}%
R_in_b^0({\bf R})=0\right. ,  \label{bscaling}
\end{eqnarray}
where $\left\langle R_i^2\right\rangle _b=\frac 1{N_b}\int d^3{\bf R}n_b^0(%
{\bf R})R_i^2$ is the average size of bosons along the $i$-axis. The last
two terms in Eq. (\ref{bscaling}), linear in $g_{bf}$, account for the
effects of boson-fermion interaction.

Analogous procedure can also be applied for the fermionic part \cite{note2}.
The equations for the scaling parameters $\gamma _i(t)$ are finally found to
take the form, 
\begin{eqnarray}
&&\ddot{\gamma}_i(t)+\omega _{if}^2(t)\gamma _i(t)-\frac{\omega _{if}^2(0)}{%
\gamma _i^3(t)}  \nonumber \\
&&+\frac{g_{bf}}{m_fN_f\left\langle R_i^2\right\rangle _f}\frac 1{\gamma
_i\prod_j\gamma _j}\int d^3{\bf R}\frac{\partial n_b^0({\bf R})}{\partial R_i%
}R_in_f^0(\frac{b_i}{\gamma _i}R_i)  \nonumber \\
&&\left. -\frac{g_{bf}}{m_fN_f\left\langle R_i^2\right\rangle _f}\frac 1{%
\gamma _i^3}\int d^3{\bf R}\frac{\partial n_b^0({\bf R})}{\partial R_i}%
R_in_f^0({\bf R})=0\right. ,  \label{fscaling}
\end{eqnarray}
where $\left\langle R_i^2\right\rangle _f=\frac 1{N_f}\int d^3{\bf R}n_f^0(%
{\bf R})R_i^2$.

The coupled set of differential equations (\ref{bscaling}) and (\ref
{fscaling}) is a generalization of Eqs. (\ref{bfree}) and (\ref{ffree}) in
the presence of the boson-fermion coupling. It determines the dynamics of
boson-fermion mixtures in the collisionless regime as far as the assumption
of the simple scaling solution is valid. We shall only be interested in the
small oscillation around the equilibrium state ($b_i,\gamma _i\approx 1$)
and apply it to study the behavior of monopole and quadrupole modes against
boson-fermion coupling. In this case, one can simplify the set of
differential equations by expanding 
\begin{eqnarray}
n_b^0(\frac{\gamma _i}{b_i}R_i) &\approx &n_b^0({\bf R})+\sum\nolimits_j%
\frac{\partial n_b^0({\bf R})}{\partial R_j}\left( \frac{\gamma _j}{b_j}%
-1\right) R_j,  \nonumber \\
n_f^0(\frac{b_i}{\gamma _i}R_i) &\approx &n_f^0({\bf R})+\sum\nolimits_j%
\frac{\partial n_f^0({\bf R})}{\partial R_j}\left( \frac{b_j}{\gamma _j}%
-1\right) R_j,  \label{expanding}
\end{eqnarray}
to obtain 
\begin{eqnarray}
&&\ddot{b}_i(t)+\omega _{ib}^2(t)b_i(t)-\frac{\omega _{ib}^2(0)}{%
b_i(t)\prod_jb_j\left( t\right) }  \nonumber \\
&&\left. +\sum\nolimits_k\frac{\omega _{ib}^2(0)B_{ik}}{b_i\prod_jb_j}\left( 
\frac{\gamma _k}{b_k}-1\right) =0\right. ,  \label{bscaling2}
\end{eqnarray}
and 
\begin{eqnarray}
&&\ddot{\gamma}_i(t)+\omega _{if}^2(t)\gamma _i(t)-\frac{\omega _{if}^2(0)}{%
\gamma _i^3(t)}+\omega _{if}^2(0)F_i\left( \frac 1{\gamma _i\prod_j\gamma _j}%
-\frac 1{\gamma _i^3}\right)  \nonumber \\
&&\left. +\sum\nolimits_k\frac{\omega _{if}^2(0)D_{ik}}{\gamma
_i\prod_j\gamma _j}\left( \frac{b_k}{\gamma _k}-1\right) =0\right. ,
\label{fscaling2}
\end{eqnarray}
where 
\begin{eqnarray*}
B_{ik} &=&\frac{g_{bf}}{m_b\omega _{ib}^2(0)N_b\left\langle
R_i^2\right\rangle _b}\int d^3{\bf R}\frac{\partial n_f^0({\bf R})}{\partial
R_i}R_iR_k\frac{\partial n_b^0({\bf R})}{\partial R_k}, \\
F_i &=&\frac{g_{bf}}{m_f\omega _{if}^2(0)N_f\left\langle R_i^2\right\rangle
_f}\int d^3{\bf R}\frac{\partial n_b^0({\bf R})}{\partial R_i}R_in_f^0({\bf R%
}), \\
D_{ik} &=&\frac{g_{bf}}{m_f\omega _{if}^2(0)N_f\left\langle
R_i^2\right\rangle _f}\int d^3{\bf R}\frac{\partial n_b^0({\bf R})}{\partial
R_i}R_iR_k\frac{\partial n_f^0({\bf R})}{\partial R_k},
\end{eqnarray*}
are the dimensionless parameters proportional to $g_{bf}$. In the case of a
cylindrical trap, those parameters can be reduced to $B_{\alpha \beta }$, $%
F_\alpha $, and $D_{\alpha \beta }$ ($\alpha ,\beta =\rho $ or $z$) in terms
of the cylindrical coordinates, whose expressions are given in Appendix.

By linearizing Eqs. (\ref{bscaling2}) and (\ref{fscaling2}) around $%
b_i,\gamma _i=1$, for each component in $\rho $ or $z$ coordinate one gets a
separated equation and thus obtains four coupled equations. The dispersion
relation for the frequency of the monopole and quadrupole modes can be
determined by the condition for existence of nontrivial solutions, that is, 
\begin{equation}
\det \left\| \omega ^2-{\bf A}_c\right\| =0,  \label{modes}
\end{equation}
where the matrix 
\[
{\bf A}_c=\left[ 
\begin{array}{cccc}
(4-B_{\rho \rho })\omega _{\bot b}^2 & (1-B_{\rho z})\omega _{\bot b}^2 & 
B_{\rho \rho }\omega _{\bot b}^2 & B_{\rho z}\omega _{\bot b}^2 \\ 
(2-B_{z\rho })\omega _{zb}^2 & (3-B_{zz})\omega _{zb}^2 & B_{z\rho }\omega
_{zb}^2 & B_{zz}\omega _{zb}^2 \\ 
D_{\rho \rho }\omega _{\bot f}^2 & D_{\rho z}\omega _{\bot f}^2 & (4-D_{\rho
\rho })\omega _{\bot f}^2 & -(F_\rho +D_{\rho z})\omega _{\bot f}^2 \\ 
D_{z\rho }\omega _{zf}^2 & D_{zz}\omega _{zf}^2 & -(2F_z+D_{z\rho })\omega
_{zf}^2 & (4+F_z-D_{zz})\omega _{zf}^2
\end{array}
\right] . 
\]
For each frequency of modes, the corresponding nontrivial solution, denoted
by ($A_{b1}$,$A_{b2}$,$A_{f1}$,$A_{f2}$), gives the amplitude of the small
density oscillations. As the system is composed of two kinds of particle,
one would expect an emergence of two type of collective oscillations for
each multipole. We thus define the mixing angle 
\begin{equation}
\theta =\arcsin \left( \sqrt{\frac{A_{b1}^2+A_{b2}^2}{%
A_{b1}^2+A_{b2}^2+A_{f1}^2+A_{f2}^2}}\right) ,  \label{mixing}
\end{equation}
to characterize the degree of mixing between bosonic and fermionic
collective motions. As a limiting case, $\theta =\pi /2$ or $0$ corresponds
to the purely (decoupled) bosonic or fermionic oscillations, respectively.

In the special case of spherical traps, the monopole and quadrupole modes
are decoupled. As shown in the Appendix, the dimensionless parameters $%
B_{\alpha \beta }$ ($F_\alpha $, $D_{\alpha \beta }$) can further be reduced
to a single value $B$ ($F,D$), 
\begin{eqnarray}
B &=&\frac{g_{bf}}{m_b\omega _b^2N_b\left\langle r^2\right\rangle _b}\int
4\pi r^2dr\frac{dn_f^0(r)}{dr}r^2\frac{dn_b^0(r)}{dr},  \nonumber \\
F &=&\frac{g_{bf}}{m_f\omega _f^2N_f\left\langle r^2\right\rangle _f}\int
4\pi r^2dr\frac{dn_b^0(r)}{dr}rn_f^0(r),  \nonumber \\
D &=&\frac{g_{bf}}{m_f\omega _f^2N_f\left\langle r^2\right\rangle _f}\int
4\pi r^2dr\frac{dn_b^0(r)}{dr}r^2\frac{dn_f^0(r)}{dr}.  \label{bfd}
\end{eqnarray}
Accordingly, the dispersion relation (\ref{modes}) for the frequency of
modes takes the following simple form, 
\begin{eqnarray}
\omega _M^2 &=&\frac 12\left\{ \left[ (5-B)\omega _b^2+(4-F-D)\omega
_f^2\right] \right.  \nonumber \\
&&\left. \pm \left( \left[ (5-B)\omega _b^2-(4-F-D)\omega _f^2\right]
^2+4BD\omega _b^2\omega _f^2\right) ^{1/2}\right\} ,  \label{mono} \\
\omega _Q^2 &=&\frac 15\left\{ \left[ (5-B)\omega _b^2+(10+5F-D)\omega
_f^2\right] \right.  \nonumber \\
&&\left. \pm \left( \left[ (5-B)\omega _b^2-(10+5F-D)\omega _f^2\right]
^2+4BD\omega _b^2\omega _f^2\right) ^{1/2}\right\} ,  \label{quad}
\end{eqnarray}
where the suffix $M$, $Q$ denote monopole and quadrupole modes, respectively.

\subsection{hydrodynamic regime}

In the hydrodynamic regime, one assumes that local equilibrium has been
established for the ground-state density profiles and it is maintained
during dynamic fluctuations of the particle densities. A useful description
of the dynamics of the Fermi gas in this regime is based on the Euler
equation of motion \cite{kagan,amoruso2}, 
\begin{eqnarray}
\frac{\partial n_f}{\partial t}+{\bf \nabla }\left( {\bf v}_fn_f\right) &=&0,
\nonumber \\
m_f\frac{\partial v_{fi}}{\partial t}+\frac 1{n_f}\frac{\partial P\left( 
{\bf r},t\right) }{\partial r_i}+m_f\sum\nolimits_jv_{fj}\frac{\partial
v_{fi}}{\partial r_j}+\frac \partial {\partial r_i}\left(
V_{ho}^f+g_{bf}n_b\right) &=&0,  \label{euler}
\end{eqnarray}
where the pressure $P\left( {\bf r},t\right) =\frac 25\frac{\hbar ^2\left(
6\pi ^2\right) ^{2/3}}{2m_f}n_f^{5/3}\left( {\bf r},t\right) $ in the local
density approximation \cite{amoruso2}. Without the boson-fermion interaction
($g_{bf}=0$), Eq. (\ref{euler}) still admits a simple scaling solution \cite
{kagan,menotti}, {\it i.e.}, 
\begin{eqnarray}
n_f\left( {\bf r},t\right) &=&\frac 1{\prod_j\gamma _j\left( t\right) }%
n_f^0\left( \frac{r_i}{\gamma _i(t)}\right) ,  \nonumber \\
v_{fi}\left( {\bf r},t\right) &=&\frac 1{\gamma _i(t)}\frac{d\gamma _i(t)}{dt%
}r_i,  \label{fansatz2}
\end{eqnarray}
which leads the following equations for the scaling parameters $\gamma _i(t)$%
: 
\begin{equation}
\ddot{\gamma}_i(t)+\omega _{if}^2(t)\gamma _i(t)-\frac{\omega _{if}^2(0)}{%
\gamma _i(t)\left[ \prod_j\gamma _j\left( t\right) \right] ^{2/3}}=0.
\label{ffree2}
\end{equation}
In the presence of a nonzero $g_{bf}$, one may still expect the validity of
such scaling ansatz in the weakly coupled limit, although there is no
verification by the other method (Note that the sum-rule approach is only
applicable in the collisionless regime). Along the same line as in the
previous subsection, after substituting the scaling ansatz (\ref{fansatz2})
into equation (\ref{euler}) and taking the moment with respect to $r_i^2$,
one ultimately finds the differential equations satisfied by $\gamma _i(t)$, 
\begin{eqnarray}
&&\ddot{\gamma}_i(t)+\omega _{if}^2(t)\gamma _i(t)-\frac{\omega _{if}^2(0)}{%
\gamma _i(t)\left[ \prod_j\gamma _j\left( t\right) \right] ^{2/3}}  \nonumber
\\
&&+\frac{g_{bf}}{m_fN_f\left\langle R_i^2\right\rangle _f}\frac 1{\gamma
_i\prod_j\gamma _j}\int d^3{\bf R}\frac{\partial n_b^0({\bf R})}{\partial R_i%
}R_in_f^0(\frac{b_i}{\gamma _i}R_i)  \nonumber \\
&&\left. -\frac{g_{bf}}{m_fN_f\left\langle R_i^2\right\rangle _f}\frac 1{%
\gamma _i\left[ \prod_j\gamma _j\right] ^{2/3}}\int d^3{\bf R}\frac{\partial
n_b^0({\bf R})}{\partial R_i}R_in_f^0({\bf R})=0\right. ,
\label{hydroscaling}
\end{eqnarray}
which differ from Eqs. (\ref{fscaling}) for the collisionless regime.

Eqs. (\ref{bscaling}) and (\ref{hydroscaling}) can be linearized around the
equilibrium state and combined to yield the determinant, 
\begin{equation}
\det \left\| \omega ^2-{\bf A}_h\right\| =0,  \label{modes2}
\end{equation}
\[
{\bf A}_h=\left[ 
\begin{array}{cccc}
(4-B_{\rho \rho })\omega _{\bot b}^2 & (1-B_{\rho z})\omega _{\bot b}^2 & 
B_{\rho \rho }\omega _{\bot b}^2 & B_{\rho z}\omega _{\bot b}^2 \\ 
(2-B_{z\rho })\omega _{zb}^2 & (3-B_{zz})\omega _{zb}^2 & B_{z\rho }\omega
_{zb}^2 & B_{zz}\omega _{zb}^2 \\ 
D_{\rho \rho }\omega _{\bot f}^2 & D_{\rho z}\omega _{\bot f}^2 & (\frac{10}3%
-\frac{2F_\rho }3-D_{\rho \rho })\omega _{\bot f}^2 & (\frac 23-\frac{F_\rho 
}3-D_{\rho z})\omega _{\bot f}^2 \\ 
D_{z\rho }\omega _{zf}^2 & D_{zz}\omega _{zf}^2 & (\frac 43-\frac{2F_z}3%
-D_{z\rho })\omega _{zf}^2 & (\frac 83-\frac{F_z}3-D_{zz})\omega _{zf}^2
\end{array}
\right] , 
\]
which gives rise to the dispersion relation in the hydrodynamic regime.

Finally, the frequency of the monopole and quadrupole modes for the
spherical traps is obtained by rewriting $B_{\alpha \beta }$ ($F_\alpha $, $%
D_{\alpha \beta }$) in terms of $B$ ($F,D$), 
\begin{eqnarray}
\omega _M^2 &=&\frac 12\left\{ \left[ (5-B)\omega _b^2+(4-F-D)\omega
_f^2\right] \right.  \nonumber \\
&&\left. \pm \left( \left[ (5-B)\omega _b^2-(4-F-D)\omega _f^2\right]
^2+4BD\omega _b^2\omega _f^2\right) ^{1/2}\right\} ,  \label{mono2} \\
\omega _Q^2 &=&\frac 15\left\{ \left[ (5-B)\omega _b^2+(5-D)\omega
_f^2\right] \right.  \nonumber \\
&&\left. \pm \left( \left[ (5-B)\omega _b^2-(5-D)\omega _f^2\right]
^2+4BD\omega _b^2\omega _f^2\right) ^{1/2}\right\} .  \label{quad2}
\end{eqnarray}
Eqs. (\ref{mono}) and (\ref{mono2}) explicitly show that the frequency of
monopole oscillations coincides in the collisionless and hydrodynamic
regime. This fact is a reminiscent of properties of a classical gas confined
in an isotropic traps ($\omega _x=\omega _y=\omega _z=\omega _0$) \cite
{boltzmann,odelin2}, in which the monopole oscillation of frequency $\omega
=2\omega _0$ is an exact undamped solution of the full Boltzmann equation.
Analogously, in the boson-fermion mixtures the monopole excitation might
also be undamped in all collisional regimes from the collisionless to the
hydrodynamic one.

At the end of this section, we briefly mention the whole process of the
numerical calculations that consists of three stages. First, one has to find
the equilibrium ground-state densities at low temperature, which {\em %
approximately} satisfy the following coupled equations in the Thomas-Fermi
approximation \cite{molmer}, 
\begin{eqnarray}
V_{ho}^b\left( \rho ,z\right) +g_{bb}n_b^0\left( \rho ,z\right)
+g_{bf}n_f^0\left( \rho ,z\right) &=&\mu _b,  \nonumber \\
\frac{\hbar ^2}{2m_f}\left( 6\pi ^2n_f^0\left( \rho ,z\right) \right)
^{2/3}+V_{ho}^f\left( \rho ,z\right) +g_{bf}n_b^0\left( \rho ,z\right)
&=&\mu _f,  \label{gs}
\end{eqnarray}
where $\mu _{b,f}$ is the chemical potential. Then one computes the
dimensionless parameters $B_{\alpha \beta }$, $F_\alpha $, and $D_{\alpha
\beta }$ ($\alpha ,\beta =\rho $ or $z$), and finally, one solves the Eqs. (%
\ref{modes}) and (\ref{modes2}) (in the case of spherical traps, one
explicitly uses Eqs. (\ref{mono}), (\ref{quad}) and (\ref{quad2})) to obtain
the frequency of the monopole and quadrupole modes. The mixing angle for
each mode is also simultaneously calculated.

\section{Result}

In this work, we have performed a numerical calculation for $N_b=N_f=10^6$,
where the number of bosons and fermions is large enough to ensure the
validity of the Thomas-Fermi approximation, {\it i.e.}, $N_ba_{bb}/a_{\bot
}^b\gg 1$ and $N_f\gg 1$ \cite{molmer,nygaard}. We take the harmonic
oscillator length $a_{\bot }^b=\sqrt{\frac \hbar {m_b\omega _{\bot b}}}$ and 
$\hbar \omega _{\bot b}$ as units, and define the scaled dimensionless
variables: the coordinates $\tilde{\rho}=\rho /a_{\bot }^b$, $\tilde{z}%
=z/a_{\bot }^b$, boson/fermion densities $\tilde{n}_{b,f}^0=n_{b,f}^0\left(
a_{\bot }^b\right) ^3$, interactions strength $\tilde{g}_{bb}=g_{bb}/\left[
\hbar \omega _{\bot b}\left( a_{\bot }^b\right) ^3\right] =4\pi
a_{bb}/a_{\bot }^b$, and chemical potentials $\tilde{\mu}_{b,f}=\mu
_{b,f}/\left( \hbar \omega _{\bot b}\right) $. We also introduce the
quantities $\alpha =m_f/m_b$, $\beta =$ $\omega _{\bot f}/\omega _{\bot b}$, 
$\lambda =\omega _{zb}/\omega _{\bot b}=\omega _{zf}/\omega _{\bot f}$, and $%
\kappa =g_{bf}/g_{bb}$ to parameterize the different mass of the two
components, anisotropy of traps, and boson-fermion coupling relative to the
boson-boson interaction. The constraint $\alpha \beta ^2=1$ is always
satisfied since in experiments both bosons and fermions experience the same
trapping potential. In the scaled units, the coupled Thomas-Fermi equations (%
\ref{gs}) take the form, 
\begin{eqnarray}
\frac 12\left( \tilde{\rho}^2+\lambda ^2\tilde{z}^2\right) +\tilde{g}%
_{bb}n_b^0+\kappa \tilde{g}_{bb}n_f^0 &=&\tilde{\mu}_b,  \nonumber \\
\frac 1{2\alpha }\left( 6\pi ^2n_f^0\right) ^{2/3}+\frac 12\left( \tilde{\rho%
}^2+\lambda ^2\tilde{z}^2\right) +\kappa \tilde{g}_{bb}n_b^0 &=&\tilde{\mu}%
_f.  \label{gs2}
\end{eqnarray}
It is convenient to obtain the solutions to Eq. (\ref{gs2}) by iterative
insertion of one density distribution in the other equation and numerically
searching for the chemical potential $\tilde{\mu}_b$ and $\tilde{\mu}_f$
yielding the desired number of particles.

We shall investigate the behavior of monopole and quadrupole modes against
boson-fermion coupling $\kappa $ for three typical values of $\tilde{g}_{bb}$
. First of all, we consider the relevant parameters for a boson-fermion
mixture composed of $^{40}${\rm K }(fermion) and{\rm \ }$^{87}${\rm Rb}
(boson), which has been recent realized by the LENS group \cite{roati}. In
order to emphasize the interplay of collective modes of bosons and fermions
due to the nonzero $g_{bf}$ (the degree of mixing will be maximum if theirs
bare mode frequencies are close to each other), we will consider the same
mass ($m_b=m_f=m$) and trapping frequency ($\omega _b=\omega _f=\omega _0$)
for bosons and fermions in most cases, although the realistic mass of $^{87}$%
{\rm Rb} is about two times larger than that of $^{40}${\rm K}. As in the
experiment, we take the radial harmonic frequency of $\omega _{\bot b}=2\pi
\times 216$ {\rm s}$^{-1}$ for $^{87}${\rm Rb} and the boson-boson $s$-wave
scattering length of $a_{bb}=110a_0=5.9$ {\rm nm}, which gives the rescaled
interaction strength $\tilde{g}_{bb}=0.1$. In the ground state the fermions
have a much broader distribution than bosons because of the Pauli principle
and the bosons are completely immersed in the Fermi sea \cite{roati}.
Secondly, as an opposite limit, we consider the case in which the fermions
and boson have approximately the same radius and significantly overlap with
each other. Within the Thomas-Fermi approximation, at $\kappa =0$, the
radius of the Bose condensate and zero-temperature Fermi gas in the scaled
units are given by $r_b=(15N_b\tilde{g}_{bb}/4\pi )^{1/5}$ and $%
r_f=(48N_f)^{1/6}$, respectively. Equating these two numbers we get the
constraint: $\tilde{g}_{bb}=2.11$. Finally, we take $\tilde{g}_{bb}=0.5$ for
the intermediate regime. It should be note that in experiments the
interaction strength $\tilde{g}_{bb}$ can be controlled by using Feshbach
resonances \cite{feshbach}.

In the section IIIA, we briefly estimate the criterion for establishing the
hydrodynamic regime. In the next sections IIIB and IIIC, we analyze the
collective modes in the collisionless and hydrodynamic regime for a
spherical trap. The results for a cylindrically symmetric trap will be
presented in section IIID.

\subsection{the criterion to establish the hydrodynamic regime}

A hydrodynamic regime is established in the low temperature alkali vapor
when the inequality 
\begin{equation}
\omega \tau \ll 1  \label{criterion}
\end{equation}
holds, where $\tau $ being the collision time for incoherent scattering of
fermions against the condensate and $\omega $ being on the scale of the trap
frequency ($\omega \simeq \omega _f$) for the low-lying modes. At low
temperature, the dominate collision procedure comes from the scattering
between a fermion and a condensate boson, which generates another fermion
and a Bogoliubov quasiparticle. In this procedure, the mean velocity of the
condensate boson is negligible relative to that of the fermions. A naive
estimate for the collision time can thus be written as 
\begin{equation}
\tau ^{-1}\approx n_b\left( 4\pi a_{bf}^2\right) v_F\left( \frac T{T_F}%
\right) ^2,  \label{time}
\end{equation}
where $n_b\left( 4\pi a_{bf}^2\right) v_F$ is the classical collisional
frequency and the factor $\left( \frac T{T_F}\right) ^2$ results from the
Pauli blocking. By setting $n_b=N_b/(4\pi r_b^3/3)$ and taking $v_F=\sqrt{%
2\mu _F/m_f}$ for a spherical trap, we approximately have 
\begin{equation}
\frac 1{\omega _f\tau }\approx 3\times 2^{1/2}\times 6^{1/6}\times \left( 
\frac{4\pi }{15\tilde{g}_{bb}}\right) ^{3/5}N_b^{2/5}N_f^{1/6}\tilde{a}%
_{bf}^2\left( \frac T{T_F}\right) ^2,  \label{criterion2}
\end{equation}
where $\tilde{a}_{bf}=a_{bf}/a_{ho}$ is the s-wave boson-fermion scattering
length in the scaled units.

For illustrative purposes we again consider the boson-fermion mixture of $%
^{40}${\rm K }and{\rm \ }$^{87}${\rm Rb} studied by the LENS group. From the
known values of the $^{40}${\rm K-}$^{87}${\rm Rb} scattering length $%
a_{bf}=300a_0$, and $\tilde{g}_{bb}=0.1$, we have 
\begin{equation}
\frac 1{\omega _f\tau }\approx 29\times \left( \frac T{T_F}\right) ^2,
\label{criterion3}
\end{equation}
for $N_b=N_f=10^6$. As anticipated earlier in the introduction, the
temperature of order $T\sim 0.5T_F$ would suffice to verify the inequality (%
\ref{criterion}) with $\omega \simeq \omega _f$, and therefore to reach the
hydrodynamic regime.

In the next subsections, we shall consider the behavior of the collective
modes against the boson-fermion interaction strength, rather than the
temperature. It should be reminded that for a fixed temperature, our results
for the hydrodynamic regime (or the collisionless regime ) are only valid at 
$\left| g_{bf}\right| \gg g_{bf}^c$ (or $\left| g_{bf}\right| \ll g_{bf}^c$%
), where $g_{bf}^c$ can be roughly determined from Eq. (\ref{criterion2}).

\subsection{collisionless modes in a spherical trap}

Figures 1 and 2, respectively, show the frequencies and mixing angles of the
monopole and quadrupole modes as a function of $\kappa $ for the case in
which the bosons and fermions have the same mass ($m_b=m_f=m$) and trapping
frequency ($\omega _b=\omega _f=\omega _0$). As we have anticipated in the
last section, there are two types of collective oscillations for each
multipole. For clarity, we plot the lower and higher frequency modes by the
solid and dashed lines, respectively.

The collisionless collective modes for the weak boson-boson interaction has
been investigated earlier by a sum-rule approach in Ref. \cite{japan00}. As
shown in the figures 1a and 2a, the result for $\tilde{g}_{bb}=0.1$ agrees
well with that obtained by the sum-rule approach in the whole regime of $%
\kappa $ if we use the same parameters (see, for example, the figures 2a and
2b in Ref. \cite{japan00}). This excellent agreement in some sense justifies
our assumption of the scaling form of solutions to Eqs. (\ref{stringari})
and (\ref{boltzmann}). We believe the same is true in the hydrodynamic
regime.

The most remarkable feature in figures 1 and 2 is the existence of a
specific value $\kappa _c\neq 0$ , at which the collective oscillation of
each mode becomes purely bosonic or fermionic. $\kappa _c$ coincides with
the critical values of phase separation for the strong boson-boson
interaction and becomes unity for the weak or medium boson-boson
interaction. The existence of $\kappa _c$ can be readily understood from the
distribution of boson and fermion densities. As shown by M\o lmer \cite
{molmer}, for a positive boson-fermion interaction, the fermions are
squeezed out the center. As $\kappa $ increases, they will eventually form a
shell-like distribution around the surface of bosons for $\kappa \geqslant 1$
and will be completely pushed away from the center at a critical value where
the phase separation occurs. For the weak or medium boson-boson interaction,
as shown in the figure 3a, precisely at $\kappa =1$ the fermions experience
a constant potential in the region occupied by bosons and therefore
uniformly distributed there. As a result, the parameters $B$ and $D$ defined
in Eqs. (\ref{bfd}) will vanish and consequently the bosonic and fermion
part in the determinant (\ref{modes}) will be completely decoupled.
Therefore the collective oscillation of modes will be purely bosonic or
fermionic. Analogous mechanics works for the case of strong boson-boson
interaction, where $B$ and $D$ will be zero at the critical value of phase
separation that is smaller than unity. Note, however, that in this case our
assumption of the scaling ansatz will apparently break down once $\kappa
>\kappa _c$, and the phase separation will lead to a decoupling of the
collective modes.

An immediate application of the above observation is that we can derive an
analytic expression for the frequency of each mode in the weak or moderately
strong boson-boson interaction in which $\kappa _c=1$. As illustrated in
figure 3b, for small values of $\kappa $, the $B$, $F$, and $D$ can be well
approximated by $b_0\kappa (1-\kappa )$, $f_0\kappa $, and $d_0\kappa
(1-\kappa )$, respectively, where $b_0=\left( dB/d\kappa \right) _{\kappa
=0},$ $f_0=\left( dF/d\kappa \right) _{\kappa =0},$ and $d_0=\left(
dD/d\kappa \right) _{\kappa =0}$. The form of $B$ and $D$ follows the fact
that they have to vanish at both $\kappa =0$ and $\kappa =1$. In the
Thomas-Fermi approximation, one may obtain, 
\begin{eqnarray}
b_0 &=&+\frac{224}\pi \left( \frac{N_f}{N_b}\right)
x^5\int\limits_0^1y^6\left( 1-x^2y^2\right) ^{1/2}dy,  \nonumber \\
f_0 &=&-\frac{256}{3\pi }x^5\int\limits_0^1y^4\left( 1-x^2y^2\right)
^{3/2}dy,  \nonumber \\
d_0 &=&+\frac{256}\pi x^7\int\limits_0^1y^6\left( 1-x^2y^2\right) ^{1/2}dy,
\label{bfd0}
\end{eqnarray}
where $x=\frac{r_b}{r_f}=\left( \frac{m_f\omega _f}{m_b\omega _b}\right)
^{1/2}(15N_b\tilde{g}_{bb}/4\pi )^{1/5}/(48N_f)^{1/6}\leqslant 1$ is the
ratio of the radius of the Bose condensate and zero-temperature Fermi gas.
The frequencies obtained by combining Eqs. (\ref{mono}), (\ref{quad}) and (%
\ref{bfd0}) are plotted in figures 1 and 2 by thin lines. We find that it is
in a good agreement with the full numerical calculations for a wide regime
of $\kappa $.

Below we discuss the behavior of the frequencies $\omega _\alpha $ of each
mode by defining three regions of $\kappa $: (I) $\kappa <0$, (II) $%
0\leqslant \kappa <\kappa _c$, and (III) $\kappa _c\leqslant \kappa $.

{\it (a) Monopole}. For a noninteracting boson-fermion system the low-lying
monopole mode is the fermionic oscillation with frequency $\omega
_M^L=2\omega _0$, while the higher mode is the bosonic one with $\omega _M^H=%
\sqrt{5}\omega _0$ in the Thomas-Fermi approximation. Around $\kappa =0$,
one may obtain $\omega _M^L\approx 2\omega _0\left( 1-\frac{d_0+f_0}8\kappa
\right) $ and $\omega _M^H\approx \sqrt{5}\omega _0\left( 1-\frac{b_0}{10}%
\kappa \right) $. The essential features of the monopole mode as a function
of $\kappa $ can be summarized as follows. ({\it i}) The curve for the
high-lying bosonic mode seems like a parabola. Indeed, the frequency for the
bosonic mode is always found to be degenerate at $\kappa =0$ and $\kappa
=\kappa _c$. In region II, the frequency varies slowly against $\kappa $ and
is slightly smaller than the value of $\sqrt{5}\omega _0$ in the
noninteracting limit. Here, the boson density distribution expands slightly
compared with that of the noninteracting boson-fermion system due to the
weakly repulsive boson-fermion interaction and simultaneously the bosons
experience a weaker effective confinement. In the region I and III, the
situation is quite different: the bosons are heavily compressed by either
the attractive or strong repulsive boson-fermion interaction. This strong
confinement leads to a steep rise of frequency with increasing $\left|
\kappa \right| $. ({\it ii}) The behavior of the frequency for low-lying
fermionic mode against $\kappa $ is more complicated. As the boson-boson
interaction strength $\tilde{g}_{bb}$ increases, the sign of the derivative
of the curve at $\kappa =0$ changes from positive to negative. At $\tilde{g}%
_{bb}=2.11$ a large dip appears in the region II. On the other hand, at a
large negative values of $\kappa $, as pointed out by Miyakawa {\it et al.} 
\cite{japan00}, we find a sharp decrease of the frequency towards the
instability point of the ground state. ({\it iii}) Finally, at large values
of $\left| \kappa \right| $ the mixing angle for both the low-lying and
high-lying modes becomes close to $\pi /4$, suggesting that the bosons and
fermions are highly correlated in the collective oscillations. The degree of
mixing is enhanced as one increases $\tilde{g}_{bb}$.

{\it (b) Quadrupole}. For the quadrupole excitation (figure 2), the lower
and higher energy mode becomes bosonic and fermionic oscillation,
respectively. To the first order of $\kappa $ the frequencies of the lower
and the higher quadrupole modes are given by $\omega _Q^L\approx \sqrt{2}%
\omega _0\left( 1-\frac{b_0}{10}\kappa \right) $ and $\omega _Q^H\approx
2\omega _0\left( 1+\frac{5f_0-d_0}{20}\kappa \right) $. For those modes,
similar mechanisms as for the monopole mode are still at work concerning the
dependence on $\kappa $. However, the role of the boson-fermion interaction
is much reduced compared with the monopole case as seen by the factor $1/5$
in Eq. (\ref{quad}), which reflects that the quadrupole oscillation has 5
different components \cite{japan00}. In addition, the behavior of the
high-lying fermionic mode is somewhat simpler. The frequency is always
decreases around $\kappa =0$ as $\kappa $ increases.

\subsection{hydrodynamic modes in a spherical trap}

In section II, we have analytically showed that in a spherical trap the
frequency of monopole oscillations coincides in the collisionless and
hydrodynamic regime. Here we find a dramatic difference for the quadrupole
mode. In the figure 4, we plot the frequencies and mixing angles of the
quadrupole modes in the hydrodynamic regime against $\kappa $. The
parameters are the same as that in figures 1 and 2. Compared with the result
for the quadrupole mode in the collisionless regime (figure 2), an
interesting feature emerges: The frequency of the low-lying mode (high-lying
mode) is always fixed to $\sqrt{2}\omega _0$ in the region I and III (region
II), independently of the value of $\kappa $, and the corresponding mixing
angle is exactly $\pi /4$ even around $\kappa =0$. This strongly suggests
that in this case the collective oscillation with equal bosonic and
fermionic amplitudes generates an exact eigenstate of the system, regardless
of the boson-fermion interaction. It resembles the Kohn mode in the
isotropic harmonic traps. Indeed, the behavior of the frequencies shown in
figure 4 is quite similar to that of a dipole mode in the collisionless
limit (see, for example, the figure 2c in Ref\cite{japan00}). The above
feature can also be explained explicitly from Eq. (\ref{quad2}). For the
case considered here, $\omega _b=\omega _f=\omega _0$, we have $\omega _Q=%
\sqrt{2}\omega _0\left\{ 1-\left[ (B+D)\pm \left| B+D\right| \right]
/10\right\} ^{1/2}$, and thus one branch of $\omega _Q$ will always be $%
\sqrt{2}\omega _0$.

The strong mixing of the bosonic and fermionic oscillation stated above in
fact arises from the ``on resonance'' condition, that is, the frequency of
the bosonic and fermionic quadrupole modes in hydrodynamic regime is
degenerate in the noninteracting limit of $\kappa =0$. As shown in figure 5,
once one moves away from the ``on resonance'' condition by changing $m_f/m_b$%
, the degree of mixing will be much reduced.

\subsection{a cylindrically symmetric trap}

In this subsection, we consider a cylindrical trap that is more relevant to
the experiment. Figures 6 and 7, respectively, display the frequencies of
each mode for a cigar-shaped ($\lambda =0.5$) and disk-shaped ($\lambda =2.0$%
) trap in the collisionless (thick lines) and hydrodynamic (thin lines)
regime. As the monopole and quadrupole modes are coupled away from $\lambda
=1$, we denote the two higher and two lower modes as quasi-monopole and
quasi-quadrupole ones, respectively. In the literature, the former one is
also called as transverse breathing mode in the limit of $\lambda
\rightarrow 0$. As one may expect, in the cylindrical trap the frequency of
quasi-monopole mode in the collisionless and hydrodynamic regime is no
longer degenerate.

{\it (a) a cigar-shaped trap}. For the quasi-monopole excitation (figure
6a), the lower and higher energy mode in both regimes are fermionic and
bosonic oscillations. In region I and III, the frequencies for the
high-lying mode in the collisionless and hydrodynamic regime are almost the
same and they only differ sightly in region II. In contrast, for the
low-lying mode the frequencies in two regimes have significant differences.
In particular, as $\kappa $ decreases towards the instability point of the
ground state, the frequency in the collisionless regime rises up steeply,
while the one in the hydrodynamic regime shows a sharp decrease. For the
quasi-quadrupole excitation (figure 6b), on the other hand, the frequencies
for each mode bears a lot of similarity as that in the spherical case. We
thus don't discuss them further.

{\it (b) a disk-shaped trap}. In this case (figure 7), the lower and higher
energy mode for both quasi-monopole and quasi-quadrupole excitations in the
collisionless regime are bosonic and fermionic oscillations, respectively.
The opposite is true in the hydrodynamic regime. Most interestingly, the
soften of the mode frequency towards the instability point of the ground
state now appears in the low-lying mode for the quasi-quadrupole
excitations, rather than the monopole one.

\section{Summary and discussion}

Ultracold boson-fermion mixtures of alkali atoms have recently been the
subject of intensive experimental research. As an important tool to
characterize the behaviour of this kind of many-body system, the
investigation of collective oscillations will be of particular interest. In
this paper, with the help of Thomas-Fermi approximation and a scaling
solution we have studied the behavior of monopole and quadrupole excitations
against the boson-fermion interaction in two limiting cases: the
collisionless and hydrodynamic regime. For a spherical trap, the frequency
of monopole mode is identical in both regimes, analogous to that of a
classical gas, which is undamped in all collisional regimes. In contrast,
the frequency of quadrupole mode differs largely in these two limits. Most
interestingly, in the case of same trapping frequency for the two components
($\omega _b=\omega _f$) and hydrodynamic regime, the quadrupole oscillations
with equal bosonic and fermionic amplitudes are found to generate an exact
eigenstate of the system, regardless of the boson-fermion interaction. It
indeed resembles the Kohn mode for the dipole excitation.

While we have restricted the discussion to the collisionless and
hydrodynamic regime, it should be explicitly remarked from the experimental
point of view that it is more interesting to investigate the {\em crossover}
between these two limits \cite{jin}, which might be realized in the
experiment by changing the temperature or controlling the particle-particle
interaction through Feshbach resonances. In theory, such crossover might be
investigated by adding a collision term ($I_{coll}$) to Eq. (\ref{boltzmann}%
).

We are aware of that the above results are based on the assumption of the
simple scaling solution. Its validity has been partly justified by the good
agreement between our results and that obtained by a sum-rule approach.
However, in case of large boson-fermion interaction, the spectrum of lowest
collective excitations might be fragmented \cite{capuzzi,japan02,bruun2}. As
a result, both the sum-rule approach and the approximation of simple scaling
solution will break down. In those regions, a refined treatment is deserved.

\section{Acknowledgements}

We acknowledge stimulating discussions with G. Modugno, M. Modugno, F.
Ferlaino, and M. Inguscio. One of us (X.-J. Liu) wishes to thank the Abdus
Salam International Centre for Theoretical Physics (ICTP) for their
hospitality during the early stages of this work. X.-J. Liu is supported by
the CAS K.C.Wang Post-doctoral Research Award Fund, the Chinese
Post-doctoral Fund and the NSF-China.

\section{Appendix}

This appendix is devoted to simplify the expressions of $B_{ik}$, $F_i$, and 
$D_{ik}$ in case of cylindrical symmetric or spherical traps. For a
cylindrical trap, the boson and fermion density distribution depend on $\rho
=\left( R_x^2+R_y^2\right) ^{1/2}$, $z=R_z$ only. By writing 
\begin{eqnarray}
\frac \partial {\partial R_x} &=&\frac{R_x}\rho \frac \partial {\partial
\rho }=\cos \phi \frac \partial {\partial \rho },  \nonumber \\
\frac \partial {\partial R_y} &=&\frac{R_y}\rho \frac \partial {\partial
\rho }=\sin \phi \frac \partial {\partial \rho },  \nonumber \\
\frac \partial {\partial R_z} &=&\frac \partial {\partial z},  \eqnum{A1}
\end{eqnarray}
one has 
\begin{eqnarray*}
B_{xx} &=&\frac{g_{bf}}{m_b\omega _{\perp b}^2(0)N_b\left\langle
R_x^2\right\rangle _b}\int d^3{\bf R}\frac{\partial n_f^0({\bf R})}{\partial
R_x}R_xR_x\frac{\partial n_b^0({\bf R})}{\partial R_x}, \\
&=&\frac{\int\limits_0^{2\pi }d\phi \cos ^4\phi }{\int\limits_0^{2\pi }d\phi
\cos ^2\phi }\left[ \frac{g_{bf}}{m_b\omega _{\perp b}^2(0)N_b\left\langle
\rho ^2\right\rangle _b}\int \rho d\rho dz\frac{\partial n_f^0}{\partial
\rho }\rho ^2\frac{\partial n_b^0}{\partial \rho }\right] \\
&=&\frac 34B_{\rho \rho },
\end{eqnarray*}
where 
\[
B_{\rho \rho }=\frac{g_{bf}}{m_b\omega _{\perp b}^2(0)N_b\left\langle \rho
^2\right\rangle _b}\int \rho d\rho dz\frac{\partial n_f^0}{\partial \rho }%
\rho ^2\frac{\partial n_b^0}{\partial \rho }. 
\]
Similarly, one finds the relations: 
\begin{equation}
\begin{array}{lll}
B_{xx}=3B_{\rho \rho }/4, & B_{xy}=B_{\rho \rho }/4, & B_{xz}=B_{\rho z}, \\ 
B_{yx}=B_{\rho \rho }/4, & B_{yy}=3B_{\rho \rho }/4, & B_{yz}=B_{\rho z}, \\ 
B_{zx}=B_{z\rho }/2, & B_{zy}=B_{z\rho }/2, & B_{zz}=B_{zz}, \\ 
&  &  \\ 
F_x=F_\rho , & F_y=F_\rho , & F_z=F_z, \\ 
&  &  \\ 
D_{xx}=3D_{\rho \rho }/4, & D_{xy}=D_{\rho \rho }/4, & D_{xz}=D_{\rho z}, \\ 
D_{yx}=B_{\rho \rho }/4, & D_{yy}=3B_{\rho \rho }/4, & D_{yz}=D_{\rho z}, \\ 
D_{zx}=D_{z\rho }/2, & D_{zy}=D_{z\rho }/2, & D_{zz}=D_{zz},
\end{array}
\eqnum{A2}
\end{equation}
where $B_{\alpha \beta }$, $F_\alpha $, and $D_{\alpha \beta }$ ($\alpha
,\beta =\rho $ or $z$) take the form, 
\begin{eqnarray}
B_{\rho \rho } &=&\frac{g_{bf}}{m_b\omega _{\perp b}^2(0)N_b\left\langle
\rho ^2\right\rangle _b}\int \rho d\rho dz\frac{\partial n_f^0}{\partial
\rho }\rho ^2\frac{\partial n_b^0}{\partial \rho },  \nonumber \\
B_{\rho z} &=&\frac{g_{bf}}{m_b\omega _{\perp b}^2(0)N_b\left\langle \rho
^2\right\rangle _b}\int \rho d\rho dz\frac{\partial n_f^0}{\partial \rho }%
\rho z\frac{\partial n_b^0}{\partial z},  \nonumber \\
B_{z\rho } &=&\frac{g_{bf}}{m_b\omega _{zb}^2(0)N_b\left\langle
z^2\right\rangle _b}\int \rho d\rho dz\frac{\partial n_f^0}{\partial z}z\rho 
\frac{\partial n_b^0}{\partial \rho },  \nonumber \\
B_{zz} &=&\frac{g_{bf}}{m_b\omega _{zb}^2(0)N_b\left\langle z^2\right\rangle
_b}\int \rho d\rho dz\frac{\partial n_f^0}{\partial z}z^2\frac{\partial n_b^0%
}{\partial z},  \nonumber \\
&&  \nonumber \\
F_\rho &=&\frac{g_{bf}}{m_f\omega _{\perp f}^2(0)N_f\left\langle \rho
^2\right\rangle _f}\int \rho d\rho dz\frac{\partial n_b^0}{\partial \rho }%
\rho n_f^0,  \nonumber \\
F_z &=&\frac{g_{bf}}{m_f\omega _{zf}^2(0)N_f\left\langle z^2\right\rangle _f}%
\int \rho d\rho dz\frac{\partial n_b^0}{\partial z}zn_f^0,  \nonumber \\
&&  \nonumber \\
D_{\rho \rho } &=&\frac{g_{bf}}{m_f\omega _{\perp f}^2(0)N_f\left\langle
\rho ^2\right\rangle _f}\int \rho d\rho dz\frac{\partial n_b^0}{\partial
\rho }\rho ^2\frac{\partial n_f^0}{\partial \rho },  \nonumber \\
D_{\rho z} &=&\frac{g_{bf}}{m_f\omega _{\perp f}^2(0)N_f\left\langle \rho
^2\right\rangle _f}\int \rho d\rho dz\frac{\partial n_b^0}{\partial \rho }%
\rho z\frac{\partial n_f^0}{\partial z},  \nonumber \\
D_{z\rho } &=&\frac{g_{bf}}{m_f\omega _{zf}^2(0)N_f\left\langle
z^2\right\rangle _f}\int \rho d\rho dz\frac{\partial n_b^0}{\partial z}z\rho 
\frac{\partial n_f^0}{\partial \rho },  \nonumber \\
D_{zz} &=&\frac{g_{bf}}{m_f\omega _{zf}^2(0)N_f\left\langle z^2\right\rangle
_f}\int \rho d\rho dz\frac{\partial n_b^0}{\partial z}z^2\frac{\partial n_f^0%
}{\partial z}.  \eqnum{A3}
\end{eqnarray}
For a spherical trap, the density distribution is a function of $r=\left(
\rho ^2+z^2\right) ^{1/2}$ only. We thus define $\rho =r\sin \theta $ and $%
z=r\cos \theta $, where $\theta \in \left[ 0,\pi \right] $. After some
straightforward algebra one finds, 
\begin{equation}
\begin{array}{llll}
B_{\rho \rho }=\frac 45B, & B_{\rho z}=\frac 15B, & B_{z\rho }=\frac 25B, & 
B_{zz}=\frac 35B, \\ 
&  &  &  \\ 
& F_\rho =F, & F_z=F, &  \\ 
&  &  &  \\ 
D_{\rho \rho }=\frac 45D, & D_{\rho z}=\frac 15D, & D_{z\rho }=\frac 25D, & 
D_{zz}=\frac 35D,
\end{array}
\eqnum{A4}
\end{equation}
where $B$, $F$ and $D$ are given by Eqs. (\ref{bfd}).

\begin{center}
{\bf Figures Captions}
\end{center}

Fig. 1. The frequencies (upper part) and mixing angles (lower part) of the
monopole mode in the collisionless regime as a function of the boson-fermion
interaction, for a spherical trap with (a) $\tilde{g}_{bb}=0.1$, (b) $0.5$
and (c) $2.11$. The other parameters are $N_b=N_f=10^6$, $m_b=m_f=m$ (or $%
\omega _b=\omega _f=\omega _0$). The low-lying and high-lying frequency mode
are denoted by the thick solid and dashed lines, respectively. The analytic
solutions obtained by combining Eqs. (\ref{mono}), (\ref{quad}) and (\ref
{bfd0}) are also plotted in figures (a) and (b) by thin lines. Note that in
figure (c), the result is not meaningful at $\kappa >\kappa _c\approx 0.69$,
where the phase separation will occur.\newline

Fig. 2. The same as in fig. 1, but for the quadrupole mode.\newline

Fig. 3. (a) The boson (solid line) and fermion (dashed line) density
distribution for a spherical trap with $\tilde{g}_{bb}=0.5$ and $\kappa =1$.
In this case, the fermions experience a constant potential in the region
occupied by bosons and therefore uniformly distributed there. (b) $%
b=B/\kappa $, $f=F/\kappa $, and $d=D/\kappa $ as a function the
boson-fermion interaction for a spherical trap with $\tilde{g}_{bb}=0.5$.
Note that precisely at $\kappa =1$, $B=D=0.$\newline

Fig. 4. The same as in fig.1, but for the quadrupole mode in the
hydrodynamic regime.\newline

Fig. 5. The frequencies (upper part) and mixing angles (lower part) of the
quadrupole mode in the hydrodynamic regime against the boson-fermion
interaction, for a spherical trap with (a) $m_f/m_b=0.8$ and (b) $%
m_f/m_b=1.03$ at $\tilde{g}_{bb}=0.5$. Since the boson and fermion trapping
frequency is not the same, the mode frequency for bosons and fermions is no
longer degenerate at $\kappa =0$, {\it i.e.}, it is out of the resonance. As
a result, the degree of mixing between bosonic and fermion collective
oscillations is much reduced.\newline

Fig. 6. The frequencies of each mode as a function of $\kappa $ for a
cigar-shaped trap with $\lambda =0.5$. (a) quasi-monopole and (b)
quasi-quadrupole. The results in the collisionless and hydrodynamic regime,
respectively, are displayed by the thick and thin lines. The other
parameters are $N_b=N_f=10^6$, $m_b=m_f=m$ and $\tilde{g}_{bb}=0.5$.\newline

Fig. 7. The same as in fig. 6, but for a disk-shaped trap with $\lambda =2.0$%
.


\begin{references}
\bibitem{note}  Of course, in the case of cylindrical traps, the anisotropy
introduces a coupling between the monopole and quadrupole modes, which are
described by the quantum numbers ($n=1$, $l=0$, $m=0$) and ($n=0$, $l=2$, $%
m=0$), respectively.
\bibitem{remark}  At low temperature, the collision term is mainly
determined by the collisions between one fermion and one condensated boson,
which scatters the fermion into another state and produce a Bogoliubov
quasiparticle. For such collision procedure, one may derive the
semi-classical equation of motions for the Wigner operator of either
fermions or Bogoliubov quasiparticles, and one may find that the
contribution of the collisions to the condensate is much smaller than that
to the fermions.
\bibitem{note2}  In this case, the average is done by multiplying the
equations by $R_i$ and $V_i$ and integrating over the phase space. See Ref. 
\cite{odelin} for details.
\end{references}
\end{document}